\documentclass[
twocolumn,
amsmath,amssymb,
prl,
raggedbottom
]{revtex4}

\usepackage{microtype}
\usepackage{graphicx, physics}
\usepackage{dcolumn}
\usepackage{bm}
\usepackage{comment}
\usepackage{amsthm}
\usepackage{mathtools}
\usepackage[dvipsnames]{xcolor}
\usepackage{epsfig}
\usepackage{dcolumn}
\usepackage{tikz}
\usepackage{braket}
\usepackage[export]{adjustbox}
\usepackage{hyperref}
\usepackage{verbatim}
\hypersetup{
	colorlinks=true,
	urlcolor=RoyalBlue,
	linkcolor=orange!80!black,
	citecolor=RoyalBlue,
	pdftitle={Unitarity Flow Conjecture},
	pdfauthor={},
	pdfdisplaydoctitle=true,
	pdfstartview=FitH
}
\usepackage{orcidlink}
\usepackage{bbm}
\usepackage{enumitem}
\usepackage[normalem]{ulem}

\def\equationautorefname~#1\null{(#1)\null}

\usetikzlibrary{calc}
\usetikzlibrary{decorations.pathmorphing, patterns, decorations.pathreplacing}
\usetikzlibrary{matrix}

\definecolor{color2}{rgb}{0.368417, 0.506779, 0.709798}
\definecolor{color3}{rgb}{0.880722, 0.611041, 0.142051}
\definecolor{color5}{rgb}{0.560181, 0.691569, 0.194885}
\definecolor{color1}{rgb}{0.922526, 0.385626, 0.209179}
\definecolor{color6}{rgb}{0.528488, 0.470624, 0.701351}
\definecolor{color4}{rgb}{0.772079, 0.431554, 0.102387}

\renewcommand{\Im}{\mathrm{Im}}
\def \be {\begin{equation}}
\def \ee {\end{equation}}

\def \d {\mathrm{d}}
\def \leq {\leqslant}

\def \M {\mathcal{M}}

\def \la {\langle}
\def \ra {\rangle}

\begin{document}

\title{Unitarity Flow Conjecture: \\
An On-shell Approach to the Renormalization Group}

\author{Ameya Chavda \orcidlink{0000-0002-1173-1605}}%
\affiliation{%
Center for Theoretical Physics, Department of Physics,\\
Columbia University, Pupin Hall,
538 West 120th Street, New York, NY 10027, USA
}

\author{Daniel McLoughlin \orcidlink{0009-0005-3535-2334}}%
\affiliation{%
Center for Theoretical Physics, Department of Physics,\\
Columbia University, Pupin Hall,
538 West 120th Street, New York, NY 10027, USA
}

\author{Sebastian Mizera \orcidlink{0000-0002-8066-5891}}%
\affiliation{%
Center for Theoretical Physics, Department of Physics,\\
Columbia University, Pupin Hall,
538 West 120th Street, New York, NY 10027, USA
}

\author{John Staunton \orcidlink{0009-0004-1661-9577}\vspace{1em}}%
\affiliation{%
Center for Theoretical Physics, Department of Physics,\\
Columbia University, Pupin Hall,
538 West 120th Street, New York, NY 10027, USA
}

\begin{abstract}
We propose that the broad architecture of the renormalization group flow in quantum field theories is, at least in part, fixed by unitarity. The precise statement is summarized in the \emph{Unitarity Flow Conjecture}, which states that the non-linear $S$-matrix identities obtained by imposing unitarity imply those needed to derive the renormalization group equations. As a proof of principle, we verify this conjecture to all loops at the leading and subleading logarithmic order in the four-dimensional massless $\lambda\phi^4$ theory using on-shell techniques, without reference to any counterterms or Feynman diagrams.
\end{abstract}

\maketitle

\section{Introduction}
Renormalization sits at the heart of the modern understanding of physics, allowing one to investigate how properties of a given physical system change depending on the energy scale $\mu$ at which it is measured. 
In the language of field theory, this structure is summarized in the framework of the renormalization group (RG).
It is difficult to overstate the ubiquity of RG; its applications range from statistical \cite{Cardy:1996} and condensed-matter systems \cite{Zinn-Justin:2007uvz}, through fluid mechanics \cite{adzhemyan1999field} and cosmology \cite{Mukhanov:2005sc}, to nuclear and particle physics \cite{Collins:1984xc}.
RG can be summarized by the equation
\be\label{eq:Callan--Symanzik}
\mu \frac{\d}{\d \mu} \mel{\text{out}}{\hat{S}}{\text{in}}  = 0\, ,\tag{R}
\ee
where $\hat{S}$ is the $S$-matrix operator and $\ket{\text{in}}$ and $\ket{\text{out}}$ are asymptotic multi-particle states. In words, \autoref{eq:Callan--Symanzik} says that physics at different scales $\mu$ stays the same, as long as the couplings and masses also change as a function of $\mu$, referred to as an \textit{RG flow}. Expanding the total $\mu$ derivative in \autoref{eq:Callan--Symanzik}, one arrives at the famous Callan--Symanzik (CS) or RG equation \cite{Callan:1970yg,Symanzik:1970rt, Symanzik:1971}, which will be made more precise in the following section.

High-precision predictions for RG flows are typically made in perturbation theory, see, e.g., \cite{Kompaniets:2017yct, Kompaniets:2020, Dupuis:2020fhh, Schnetz:2022nsc, Bednyakov:2025, Borinsky:2025, Kovyrshin:2025, Huang:2025, Gracey:2025, Schnetz:2025} for cutting-edge results. From this perspective, it is highly non-trivial that renormalized perturbation theory works at all beyond one-loop order, as it relies on an intricate pattern of relations between Feynman integrals at different perturbative orders. Nevertheless, a celebrated theorem due to Bogoliubov, Parasiuk, Hepp, and Zimmermann \cite{Bogoliubov:1957gp,Hepp:1966eg,Zimmermann:1969jj} proves that RG is self-consistent as a consequence of a web of non-linear identities between different matrix elements.

There is, however, another source of non-linear identities between matrix elements: unitarity. It encodes the physical principle of probability conservation and in equations can be written as the operator statement
\be\label{eq:unitarity}
\hat{S} \hat{S}^\dagger = \mathbbm{1}\, ,\tag{U}
\ee
which simply says that probabilities have to sum to one. The goal of this letter is to put forward a conjecture that \autoref{eq:Callan--Symanzik} is, at least in part, a consequence of \autoref{eq:unitarity}.

Over the years, several similarities between \autoref{eq:Callan--Symanzik} and \autoref{eq:unitarity} have been observed. For example, Koschinski, Polyakov, and Vladimirov \cite{Koschinski:2010} found that leading divergences of 4-particle amplitudes are fixed as a consequence of unitarity, analyticity, and permutation symmetry. More recently, Caron-Huot and Wilhelm \cite{Caron-Huot:2016} used a similar construction to compute $\beta$ functions in QCD and Yukawa theory through two loops. These investigations were extended to rapidity RG \cite{Rothstein:2023dgb} and applied to the Standard Model Effective Field Theory \cite{EliasMiro:2020tdv,Baratella:2020lzz,Jiang:2020mhe,Bern:2020}, among others \cite{Baratella:2022, Bresciani:2023, DeAngelis:2023, Bresciani:2024, Mo:2025, Aebischer:2025}. Broadly speaking, the intuition behind these developments is that leading divergences involve logarithms of the form $\log \big(\tfrac{\mu^2}{p^2} \big)$ and changing the energy scale $\mu$ can be equivalently written as computing the imaginary part:
\be
\frac{2}{\pi}\, \Im \log\left(\frac{\mu^2}{p^2}\right)
=
\mu \frac{\d \log\left(\frac{\mu^2}{p^2}\right)}{\d\mu}
\ee
for timelike momenta, $p^2 < 0$. The imaginary part is then related to the long-range propagation of on-shell states through the optical theorem, see, e.g., \cite{Itzykson:1980,Britto:2024}.

We extend this intuition beyond leading logarithms and formulate the \emph{Unitarity Flow Conjecture} (UFC), which schematically states that
\be
\text{\autoref{eq:unitarity}} \subseteq \text{\autoref{eq:Callan--Symanzik}}
\ee
and will be made more precise in the following sections. In words, the non-linear relations derived from unitarity imply, at least in part, the structure of the CS equation. The converse statement is certainly not true, since there exist non-unitary theories that have RG flows, e.g., in dissipative systems \cite{Itzykson:1986}.
In the weak form, the UFC states \autoref{eq:unitarity} $\subset$ \autoref{eq:Callan--Symanzik}, that a part of RG equations can be obtained from unitarity. In the strong form, the UFC states \autoref{eq:unitarity} $=$ \autoref{eq:Callan--Symanzik}, that RG and unitarity have the same content.

In this letter, we provide evidence for the strong UFC in the case of the massless $\lambda\phi^4$ theory in four dimensions to all loops through the subleading logarithmic order. This theory is enough to illustrate the non-trivial aspects of the UFC without unnecessary technical complications. We will show that the information content of \autoref{eq:Callan--Symanzik} at each logarithmic order can be derived from a series of unitarity cuts. These cuts, in turn, only require knowledge of the on-shell amplitudes.
 
The UFC is a part of a broader program to give an on-shell understanding of renormalization, see, e.g., \cite{Koschinski:2010,Huang:2012aq, Cheung:2015, Caron-Huot:2016, Baratella:2021, Arkani-Hamed:2022cqe,Rothstein:2023dgb,Chala:2023jyx,Chala:2023xjy,Liao:2025npz, Freidel:2025}. The on-shell approach highlights a qualitative improvement over standard off-shell computations by avoiding swaths of redundant information present at each order in the perturbative expansion. In particular, no counterterms or Feynman diagrams will appear in this letter.

\vspace{-1em}
\section{Architecture of renormalization}

As a starting point for an on-shell theory of renormalization, we are going to use the following minimal set of axioms: (I) finite $S$-matrix elements consistent with all symmetries exist, (II) $S$-matrix elements run with an energy scale $\mu$, (III) $S$-matrix elements are unitary for every $\mu$. Let us explain how each of them is implemented in practice in the massless $\lambda \phi^4$ theory.

\textbf{(I) Symmetries.}
Since we are interested in the running of the marginal coupling $\lambda$ in four dimensions, we will consider the 4-particle scattering amplitude $i\M_{4} = \la p_3 p_4 | \hat{S} - \mathbbm{1} | p_1 p_2 \ra $, which depends on the Mandelstam invariants $s = -\left(p_1 {+} p_2\right)^2 > 0$, $t = -\left(p_2 {-} p_3\right)^2 < 0$, and $u = -\left(p_2 {-} p_4\right)^2 < 0$ with $s+t+u=0$.

We will work in the hard scattering limit, $|s| \sim |t| \sim |u| \gg \mu^2$. In this limit, the amplitude is dominated by large logarithms. Therefore, the renormalized 4-particle amplitude admits the following loop expansion:
\begin{equation}
\label{eq:Ansatz4}
\M_{4} = \sum_{L = 0}^{\infty} \frac{\left(-\lambda\right)^{L + 1}}{\left(16\pi^2\right)^L} \M_{4}^{\left(L\right)}
\end{equation}
with
\begin{equation}
\label{eq:General4pt}
\begin{aligned}
\M_{4}^{\left(L\right)} &= \sum_{0 \leq k_1, k_2, k_3 \leq L} m_{L, \left\{k_1, k_2, k_3\right\}} \\ 
&\hspace{0.5cm} \times \log^{k_1}\left(\frac{\mu^2}{-s}\right) \log^{k_2}\left(\frac{\mu^2}{-t}\right) \log^{k_3}\left(\frac{\mu^2}{-u}\right),
\end{aligned}
\end{equation}
where $m_{L, \left\{k_1, k_2, k_3\right\}}$ denotes the coefficient at $L^{\text{th}}$-loop order and powers $k_1$, $k_2$, and $k_3$ of each of the logs. At tree level, we have $m_{0, \left\{0, 0, 0\right\}} = 1$. Permutation symmetry implies that the amplitude is $\left(s, t, u\right)$-symmetric and hence the coefficient $m_{L, \left\{k_1, k_2, k_3\right\}}$ is also  $(k_1, k_2, k_3)$-symmetric. Moreover, $m_{L, \left\{k_1, k_2, k_3\right\}}$ are real and independent of the kinematics \cite{Kleinert:2001}. We will refer to the coefficients with $k_1 + k_2 + k_3 = L$ and $L-1$ as leading and subleading respectively. The ansatz in \autoref{eq:General4pt} is therefore the most general amplitude of a marginal theory with no IR divergences that has permutation symmetry, and real coefficients in the hard scattering limit. For further discussion on the IR features of this theory, see \cite[Chapter 12]{Kleinert:2001}.

The form of the 2-particle amplitude (which is related to the sum over all 1PI propagator corrections \cite[Eq. (9.25)]{Kleinert:2001}) is similar: 
\begin{equation}
\label{eq:Ansatz2}
\M_{2} = p^2 - \sum_{L = 1}^{\infty} \frac{\left(-\lambda\right)^{L + 1}}{\left(16\pi^2\right)^L} \M_{2}^{\left(L\right)}
\end{equation}
with
\begin{equation}
\M_{2}^{\left(L\right)} = p^2\sum_{k = 0}^{L} n_{L, k} \log^k\left(\frac{\mu^2}{p^2}\right).
\end{equation}
We will similarly refer to $n_{L, L}$ and $n_{L, L - 1}$ as leading and subleading coefficients, respectively. 

We note that the inclusion of constant terms in these expansions, $m_{L, \left\{0, 0, 0\right\}}$ and $n_{L, 0}$, encodes the scheme dependence of the amplitude. These are fixed by the definition of the physical parameters. For example, the on-shell scheme defines the physical value of the coupling $\lambda$ by continuing to the unphysical kinematic values $s=t=u=-\mu^2$, which sets all of these scheme-dependent terms to zero, $m_{L, \left\{0, 0, 0\right\}} = n_{L, 0} = 0$.

\textbf{(II) Running.} The S-matrix elements run with the scale $\mu$, meaning the $n$-particle amplitude $\M_n\left(\mu, \lambda\left(\mu\right)\right)$ obeys
\begin{equation}
\label{eq:CSonM}
\mu\partial_\mu \M_n = - \beta \partial_\lambda \M_n + n \gamma \M_n.
\end{equation}
By the chain rule, 
\begin{equation}
\beta\left(\lambda\right) = \mu\partial_\mu \lambda.
\end{equation}
The appearance of $\gamma$ follows from the fact that the one-particle states $\ket{p_i}$ also get renormalized, 
\begin{equation}
\gamma\left(\lambda\right) = \mu \partial_\mu \ket{p_i},
\end{equation}
with the factor of $n$ in \autoref{eq:CSonM} coming from the amplitude constructed out of $n$-particle states. We choose to write \autoref{eq:CSonM} in this way since it is the precise form of the CS equation mentioned below \autoref{eq:Callan--Symanzik}, but we stress here that $\beta$ and $\gamma$  are arbitrary functions of $\lambda$ and do not rely on counterterms. Note that, crucially, these functions do not have an explicit $\mu$-dependence, instead characterizing the RG flow of the theory. 

\begin{figure*}[t]
    \centering
\begin{equation*}
%\hspace{-1cm}
2\: \Im\left(\hspace{0.1cm}
\begin{gathered}
\begin{tikzpicture}[scale=0.6]
\draw[line width=0.3mm] (-1, 1) -- (-0.5*0.707, 0.5*0.707);
\draw[line width=0.3mm] (-1, -1) -- (-0.5*0.707, -0.5*0.707);
\filldraw[pattern=north west lines] (0, 0) circle (0.5cm);
\draw[line width=0.3mm] (0, 0) circle (0.5cm);
\draw[line width=0.3mm] (0.5*0.707, 0.5*0.707) -- (1, 1);
\draw[line width=0.3mm] (0.5*0.707, -0.5*0.707) -- (1, -1);
\end{tikzpicture}
\end{gathered}\hspace{0.1cm}\right) =
\begin{gathered}
\begin{tikzpicture}[scale=0.6, remember picture]
\draw[line width=0.3mm] (-1, 1) -- (-0.5*0.707, 0.5*0.707);
\draw[line width=0.3mm] (-1, -1) -- (-0.5*0.707, -0.5*0.707);
\filldraw[pattern = north west lines] (0, 0) circle (0.5cm);
\draw[line width=0.3mm] (0, 0) circle (0.5cm);
\draw[line width=0.3mm] (0.5*0.866, 0.5*0.5) -- (2 - 0.5*0.866, 0.5*0.5);
\draw[line width=0.3mm] (0.5*0.866, -0.5*0.5) -- (2 - 0.5*0.866, -0.5*0.5);
\filldraw[pattern=north west lines] (2, 0) circle (0.5cm);
\draw[line width=0.3mm] (2, 0) circle (0.5cm);
\draw[line width=0.3mm] (2 + 0.5*0.707, 0.5*0.707) -- (3, 1);
\draw[line width=0.3mm] (2 + 0.5*0.707, -0.5*0.707) -- (3, -1);
\filldraw[orange!10, opacity=0.5] (1, 1) rectangle (3.2, -1);
\draw[line width=0.5mm, dashed, orange] (1, 1) -- (1, -1);
% Coordinates for the first brace
\coordinate (brace1_top) at (2.75, -1.25);
\coordinate (brace1_left) at (-0.75, 1.5);
\coordinate (brace1_bottom) at (-0.75, -1.25);
\end{tikzpicture}
\end{gathered} \: \: + \: \:
\begin{gathered}
\begin{tikzpicture}[scale=0.6, remember picture] 
\draw[line width=0.3mm] (-1, 1) -- (-0.5*0.707, 0.5*0.707);
\draw[line width=0.3mm] (-1, -1) -- (-0.5*0.707, -0.5*0.707);
\draw[pattern=north west lines] (0, 0) circle (0.5cm);
\draw[line width=0.3mm] (0, 0) circle (0.5cm);
\draw[line width=0.3mm] (0.5*0.866, 0.5*0.5) -- (2 - 0.5*0.866, 0.5*0.5);
\draw[line width=0.3mm] (0.5*0.985, 0.5*0.174) -- (2 - 0.5*0.985, 0.5*0.174);
\draw[line width=0.3mm] (0.5*0.985, -0.5*0.174) -- (2 - 0.5*0.985, -0.5*0.174);
\draw[line width=0.3mm] (0.5*0.866, -0.5*0.5) -- (2 - 0.5*0.866, -0.5*0.5);
\draw[pattern = north west lines] (2, 0) circle (0.5cm);
\draw[line width=0.3mm] (2, 0) circle (0.5cm);
\draw[line width=0.3mm] (2 + 0.5*0.707, 0.5*0.707) -- (3, 1);
\draw[line width=0.3mm] (2 + 0.5*0.707, -0.5*0.707) -- (3, -1);
\filldraw[orange!10, opacity=0.5] (1, 1) rectangle (3.2, -1);
\draw[line width=0.5mm, dashed, orange] (1, 1) -- (1, -1);
% Coordinate for the bottom of the second brace
\coordinate (brace2_top) at (-0.75, 1.55);
\coordinate (brace2_bottom) at (2.75, 1.55);
\end{tikzpicture}
\end{gathered} \: \:+ \dots
\end{equation*}
\begin{tikzpicture}[overlay, remember picture]
    % Leading term brace
    \draw [decorate, decoration={brace, amplitude=6pt}] 
          (brace1_top) -- (brace1_bottom) 
          node [midway, yshift=-0.5cm] {Leading};
          
    % subleading brace
    \draw [decorate, decoration={brace, amplitude=6pt}] 
          (brace1_left) -- (brace2_bottom)
          node [midway, yshift=0.5cm] {Subleading};
\end{tikzpicture}

\vspace{0.3cm}

\begin{equation*}
\begin{aligned}
2\: \text{Im}\left(
\begin{gathered}
\begin{tikzpicture}[scale=0.6]
\draw[line width=0.3mm] (-1.5, 0) -- (-0.5, 0);
\draw[pattern= north west lines] (0, 0) circle (0.5cm);
\draw[line width=0.3mm] (0, 0) circle (0.5cm); 
\draw[line width=0.3mm] (0.5, 0) -- (1.5, 0);
\end{tikzpicture}
\end{gathered}
\right) = 
\begin{gathered}
\vspace{-0.85cm}
\begin{tikzpicture}[scale=0.6]
\draw[line width=0.3mm] (-1.5, 0) -- (-0.5, 0);
\draw[pattern= north west lines] (0, 0) circle (0.5cm);
\draw[line width=0.3mm] (0, 0) circle (0.5cm);
\draw[line width=0.3mm] (0.5*0.866, 0.5*0.5) -- (2 - 0.5*0.866, 0.5*0.5);
\draw[line width=0.3mm] (0.5*0.866, -0.5*0.5) -- (2 - 0.5*0.866, -0.5*0.5);
\draw[pattern=north west lines] (2, 0) circle (0.5cm);
\draw[line width=0.3mm] (2, 0) circle (0.5cm);
\draw[line width=0.3mm] (0.5, 0) -- (1.5, 0);
\draw[line width=0.3mm] (2.5, 0) -- (3.5, 0);
\filldraw[orange!10, opacity=0.5] (1, 1) rectangle (3.7, -1);
\draw[line width=0.5mm, dashed, orange] (1, 1) -- (1, -1);
\draw[decorate, decoration={brace, amplitude=6pt}] (3, -1.15) -- node[anchor=east, xshift=1cm, yshift= -0.5cm] {Subleading} (-1, -1.15);
\end{tikzpicture}
\end{gathered} \: \: + \dots
\end{aligned}
\end{equation*}
\vspace{0.1cm}
\caption{Diagrammatic representation of the generalized optical theorem. Terms contributing to the leading and subleading divergences of the 2- and 4-particle amplitudes are indicated above. The ellipsis contains terms at further subleading orders.}
\label{fig:unitarity}
\end{figure*}

\textbf{(III) Unitarity.} 
Unitarity can be stated in the form of the generalized optical theorem (unitarity equation),
\begin{equation}
\label{eq:OpticalTheorem}
\Im\, \M_{n_1 + n_2} = \frac{1}{2} \sum_{k\text{-cuts}} \int \M_{n_1 + k}\, \M^\ast_{n_2 + k} \dd{\Phi}_{k},
\end{equation}
where $\dd{\Phi}_{k}$ is the $k$-particle Lorentz invariant phase space measure (recall that $\M$'s do contain disconnected terms but not the identity), see, e.g., \cite{Britto:2024}. From now on, we assume $\mathbb{Z}_2$ symmetry, $\ket{p_i} \rightarrow - \ket{p_i}$ for all $i$, which implies that $\M_n$ vanishes for odd $n$. Hence, only even-particle amplitudes contribute to the sum. We will diagrammatically represent unitarity cuts putting $k$ particles on-shell (called $k$-cuts) with an orange dashed line and complex conjugation with shading, see \autoref{fig:unitarity}. Upon using the loop expansion, such as \autoref{eq:Ansatz4}, the unitarity equation recursively relates higher to lower loop coefficients.

\vspace{-1em}
\section{Unitarity Constraints}

The central observation is that out of the infinite sum in \autoref{eq:OpticalTheorem}, only a finite number of terms contribute to the leading and subleading coefficients. We will use the cuts shown in \autoref{fig:unitarity} to determine these coefficients, $m_{L, \left\{k_1, k_2, k_3\right\}}$ and $n_{L, k}$, to all loop orders, without any use of counterterms or Feynman diagrams. These relations, we argue, are enough to reconstruct the leading and subleading $\beta$ and $\gamma$ functions. Moreover, we will demonstrate that the constraints on amplitudes given by RG and unitarity differ only by the initial conditions of the recursion relations. 

\textbf{Leading 4-particle coefficients, $\bm{m_{L,L}}$.} From unitarity on the 4-particle amplitude we find that, since the sum over 2- and 4-cuts only depends on $\log \big(\tfrac{\mu^2}{s}\big)$, the coefficients of the amplitude that mix logarithms in different Mandelstam variables are all zero. In other words, unitarity tells us that only
\begin{equation}
m_{L,k} \equiv m_{L, \left\{k, 0, 0\right\}} = m_{L, \left\{0, k, 0\right\}} = m_{L, \left\{0, 0, k\right\}}
\end{equation}
are non-zero for $k = L$ and $L - 1$. For example, terms of the form $\log^{L'} (\tfrac{\mu^2}{-s}) \log^{L - L'} (\tfrac{\mu^2}{-t})$ are absent. This result is reminiscent of the Steinmann relations \cite{Steinmann1960a,Steinmann1960b}, which constrain double discontinuities in overlapping channels. The amplitude \autoref{eq:General4pt} therefore simplifies to
\begin{equation}\label{eq:Simplified4pt}
\M_{4}^{\left(L\right)} = \sum_{k = 0}^{L} m_{L, k} \bigg[\!\log^k\left(\tfrac{\mu^2}{-s}\right) + \log^k\left(\tfrac{\mu^2}{-t}\right) + \log^k\left(\tfrac{\mu^2}{-u}\right)\!\bigg].
\end{equation}
Note that when $k = 0$ we have $m_{0, 0} = 1/3$, so that the tree level 4-particle amplitude is $\M^{(0)}_{4} = -\lambda$.

The recursion relations among the leading coefficients $m_{L, L}$ are solely given by the sum over 2-cuts. In particular, matching orders in $\lambda$ we find
\begin{equation}
\Im\, \M_{4}^{(L)} = \frac{1}{2} \sum_{L' = 0}^{L - 1} \int \M_{4}^{(L')}\, \M_{4}^{\ast (L - L' - 1)} \dd{\Phi}_2 + \dots\;. 
\end{equation}
Upon plugging in the ansatz \autoref{eq:Simplified4pt} on both sides of the equation, one can match powers in leading log.

This procedure results in a recursion relation for the leading coefficients: 
\begin{equation}
m_{L, L} = \frac{9}{2L} \sum_{L' = 0}^{L - 1} m_{L', L'} m_{L - L' - 1, L - L' - 1},
\end{equation}
whose solution is
\begin{equation}
\label{eq:LL4}
m_{L, L} = \frac{1}{3}\left(\frac{3}{2}\right)^L.
\end{equation}

\textbf{Leading 2-particle coefficients, $\bm{n_{L,L}}$.}
We now turn to the 2-particle amplitude. Since 1PI diagrams, by definition, have no 1-cuts, the first contribution is from 3-cuts. Consequently, there is no leading log contribution on the cut side of the unitarity equation and
\begin{equation}
n_{L, L} = 0.
\end{equation}
The first non-zero coefficients are therefore subleading.

\textbf{Subleading 2-particle coefficients, $\bm{n_{L,L-1}}$.}
The unitarity equation relates the subleading coefficients, $n_{L, L - 1}$, on the left-hand side of the unitarity equation to products of leading 4-particle coefficients on the right-hand side,
\begin{equation}
\Im\, \M_2^{\left(L\right)} = \frac{1}{2} \sum_{L' = 0}^{L - 2} \int \M_{4}^{(L')}\, \M_{4}^{\ast (L - L' - 2)} \dd{\Phi}_3 + \dots\;.
\end{equation}
Once again, power counting shows that the higher-particle cuts will not contribute to subleading order in logs. By plugging in the results of the leading 4-particle coefficients on the right-hand side, we can constrain the 2-particle subleading coefficients to be
\begin{equation}
\label{eq:LL2}
n_{L, L - 1} = -\frac{1}{27} \left(\frac{3}{2}\right)^L. 
\end{equation}
As a consistency check, positivity of $m_{L,L}$ and negativity of $n_{L,L-1}$ are guaranteed by the optical theorem.

\begin{figure*}[t]
\begin{equation*}
2\: \text{Im}\left(\hspace{0.1cm}
\begin{gathered}
\begin{tikzpicture}[scale=0.6]
\draw[line width=0.3mm] (-1, 1) -- (-0.5*0.707, 0.5*0.707);
\draw[line width=0.3mm] (-1, -1) -- (-0.5*0.707, -0.5*0.707);
\filldraw[pattern=north west lines] (0, 0) circle (0.5cm);
\draw[line width=0.3mm] (0, 0) circle (0.5cm);
\draw[line width=0.3mm] (0.5*0.707, 0.5*0.707) -- (1, 1);
\draw[line width=0.3mm] (0.5*0.707, -0.5*0.707) -- (1, -1);
\draw[line width=0.3mm] (0.5*0.97, 0.5*0.26) -- (0.92*0.75 + 0.5, 0.38*0.75);
\draw[line width=0.3mm] (0.5*0.97, -0.5*0.26) -- (0.92*0.75 + 0.5, -0.38*0.75);
\end{tikzpicture}
\end{gathered}
\hspace{0.1cm}\right)
= \: \:
\begin{gathered}
\begin{tikzpicture}[scale=0.6]
\draw[line width=0.3mm] (-1, 1) -- (-0.5*0.707, 0.5*0.707);
\draw[line width=0.3mm] (-1, -1) -- (-0.5*0.707, -0.5*0.707);
\filldraw[pattern=north west lines] (0, 0) circle (0.5cm);
\draw[line width=0.3mm] (0, 0) circle (0.5cm);
\draw[line width=0.3mm] (0.5*0.707, 0.5*0.707) -- (3.25, 1.25);
\draw[line width=0.3mm] (0.5, 0) -- (1.5, 0);
\filldraw[pattern=north west lines] (2, 0) circle (0.5cm);
\draw[line width=0.3mm] (2, 0) circle (0.5cm);
\draw[line width=0.3mm] (2 + 0.5*0.707, 0.5*0.707) -- (3.25, 0.75);
\draw[line width=0.3mm] (2 + 0.5, 0) -- (3.25, 0);
\draw[line width=0.3mm] (2 + 0.5*0.707, -0.5*0.707) -- (3.25, -0.75);
\filldraw[orange!10, opacity=0.5] (1, 1.5) rectangle (3.5, -1.5);
\draw[line width=0.5mm, dashed, orange] (1, 1.5) -- (1, -1.5);
\end{tikzpicture}
\end{gathered} \: \: + \: \: 
\begin{gathered}
\begin{tikzpicture}[scale=0.6]
\draw[line width=0.3mm] (-1, 1) -- (-0.5*0.707, 0.5*0.707);
\draw[line width=0.3mm] (-1, -1) -- (-0.5*0.707, -0.5*0.707);
\filldraw[pattern=north west lines] (0, 0) circle (0.5cm);
\draw[line width=0.3mm] (0, 0) circle (0.5cm);
\draw[line width=0.3mm] (0.5*0.707, 0.5*0.707) -- (3.25, 0.75);
\draw[line width=0.3mm] (0.5, 0) -- (1.5, 0);
\filldraw[pattern=north west lines] (2, 0) circle (0.5cm);
\draw[line width=0.3mm] (2, 0) circle (0.5cm);
\draw[line width=1.3mm, white] (2 + 0.5*0.707, 0.5*0.707) -- (3.25, 1.25);
\draw[line width=0.3mm] (2 + 0.5*0.707, 0.5*0.707) -- (3.25, 1.25);
\draw[line width=0.3mm] (2 + 0.5, 0) -- (3.25, 0);
\draw[line width=0.3mm] (2 + 0.5*0.707, -0.5*0.707) -- (3.25, -0.75);
\filldraw[orange!10, opacity=0.5] (1, 1.5) rectangle (3.5, -1.5);
\draw[line width=0.5mm, dashed, orange] (1, 1.5) -- (1, -1.5);
\end{tikzpicture}
\end{gathered} \: \: + \: \: 
\begin{gathered}
\begin{tikzpicture}[scale=0.6, yscale=-1]
\draw[line width=0.3mm] (-1, 1) -- (-0.5*0.707, 0.5*0.707);
\draw[line width=0.3mm] (-1, -1) -- (-0.5*0.707, -0.5*0.707);
\filldraw[pattern=north west lines] (0, 0) circle (0.5cm);
\draw[line width=0.3mm] (0, 0) circle (0.5cm);
\draw[line width=0.3mm] (0.5*0.707, 0.5*0.707) -- (3.25, 0.75);
\draw[line width=0.3mm] (0.5, 0) -- (1.5, 0);
\filldraw[pattern=north west lines] (2, 0) circle (0.5cm);
\draw[line width=0.3mm] (2, 0) circle (0.5cm);
\draw[line width=1.3mm, white] (2 + 0.5*0.707, 0.5*0.707) -- (3.25, 1.25);
\draw[line width=0.3mm] (2 + 0.5*0.707, 0.5*0.707) -- (3.25, 1.25);
\draw[line width=0.3mm] (2 + 0.5, 0) -- (3.25, 0);
\draw[line width=0.3mm] (2 + 0.5*0.707, -0.5*0.707) -- (3.25, -0.75);
\filldraw[orange!10, opacity=0.5] (1, 1.5) rectangle (3.5, -1.5);
\draw[line width=0.5mm, dashed, orange] (1, 1.5) -- (1, -1.5);
\end{tikzpicture}
\end{gathered} \: \: + \: \: 
\begin{gathered}
\begin{tikzpicture}[scale=0.6, yscale=-1]
\draw[line width=0.3mm] (-1, 1) -- (-0.5*0.707, 0.5*0.707);
\draw[line width=0.3mm] (-1, -1) -- (-0.5*0.707, -0.5*0.707);
\filldraw[pattern=north west lines] (0, 0) circle (0.5cm);
\draw[line width=0.3mm] (0, 0) circle (0.5cm);
\draw[line width=0.3mm] (0.5*0.707, 0.5*0.707) -- (3.25, 1.25);
\draw[line width=0.3mm] (0.5, 0) -- (1.5, 0);
\filldraw[pattern=north west lines] (2, 0) circle (0.5cm);
\draw[line width=0.3mm] (2, 0) circle (0.5cm);
\draw[line width=0.3mm] (2 + 0.5*0.707, 0.5*0.707) -- (3.25, 0.75);
\draw[line width=0.3mm] (2 + 0.5, 0) -- (3.25, 0);
\draw[line width=0.3mm] (2 + 0.5*0.707, -0.5*0.707) -- (3.25, -0.75);
\filldraw[orange!10, opacity=0.5] (1, 1.5) rectangle (3.5, -1.5);
\draw[line width=0.5mm, dashed, orange] (1, 1.5) -- (1, -1.5);
\end{tikzpicture}
\begin{tikzpicture}[overlay, remember picture]
    % subleading brace
    \draw [decorate, decoration={brace, amplitude=6pt}] 
          (0,-0.2) -- (-13,-0.2)
          node [midway, yshift=-0.6cm] {Subleading};
\end{tikzpicture}
\end{gathered} \: \:  + \dots 
\end{equation*}
\vspace{0.5cm}
\caption{Generalized optical theorem applied to the $6$-particle amplitude. The logs of the 6-particle amplitude arise from cuts through a single internal line. Unitarity then implies that these 6-particle amplitudes are products of 4-particle amplitudes. Out of the 10 permutations, the 4 displayed above are the only ones that contribute to subleading coefficients \cite{CMMS}.}
\label{fig:6ParticleFactorization}
\end{figure*}

\textbf{Subleading 4-particle coefficients, $\bm{m_{L,L-1}}$.}
Finally, we can include both the 2-cuts and 4-cuts of \autoref{fig:unitarity} to constrain the subleading 4-particle amplitude coefficients, $m_{L, L - 1}$. However, the 4-cuts contain 6-particle amplitudes, 
\begin{equation}
\begin{aligned}
\Im\, \M_{4}^{(L)} &= \frac{1}{2} \sum_{L' = 0}^{L - 1} \int \M_{4}^{(L')}\, \M_{4}^{\ast (L - L' - 1)} \dd{\Phi}_2 \\
&\hspace{0.4cm} + \frac{1}{2} \sum_{L' = 0}^{L - 3} \M_{6}^{(L')}\, \M_{6}^{\ast(L - L' - 3)} \dd{\Phi}_4 + \dots\;. 
\end{aligned}
\end{equation}
Note that, by the same power-counting arguments as before, the order in log from the 4-particle cuts is sub-subleading as opposed to subleading. The difference in this case is that the 4-particle phase space itself will contain divergences and contribute an enhancement by an additional power in log. Using the same reasoning as for the 2- and 4-particle amplitudes, the leading contribution to the 6-particle amplitude comes from products of 4-particle amplitudes, as shown in \autoref{fig:6ParticleFactorization}. Said in words, unitarity implies the factorization of the 6-particle amplitude into products of 4-particle amplitudes connected with a propagator, hence
\begin{equation}
\M_{6} = \sum_{s_{ijk}} \M_{4} \frac{i}{s_{ijk} + i\varepsilon} \M_{4}^\ast + \dots,
\end{equation}
where the sum occurs over all ten channels and we used $\Im \frac{1}{x + i\varepsilon} = -\pi \delta(x)$. Since the 6-particle amplitude at leading order is expressible in terms of 4-particle amplitude coefficients, the unitarity equation once again gives a closed relationship among the leading and now subleading coefficients of the 4-particle amplitude. Plugging in the result for the leading coefficients gives a recursion for the subleading coefficients, 
\begin{equation}
m_{L, L - 1} = \left(\frac{3}{2}\right)^{\! L}\Bigg[\frac{17 L {+} 2}{81} + \frac{2}{L {-} 1}\sum_{L' = 1}^{L - 1} \left(\frac{2}{3}\right)^{\! L'} \!\!\!\! m_{L', L' - 1}\Bigg],
\end{equation}
whose solution is
\begin{equation}
\label{eq:SLL4}
m_{L, L - 1} = \left(\frac{3}{2}\right)^L \frac{L\left(54 m_{1, 0} - 32\right) + 34 L H_L - 2}{81}, 
\end{equation}
where $H_L$ is the $L^\mathrm{th}$ harmonic number. Recall that $m_{1, 0}$ is scheme-dependent and sets a boundary condition for the recursion.

\textbf{Consistency checks.}
There are several checks for these results. We explicitly computed the renormalized amplitude up to 3-loop order in renormalized perturbation theory, where 4-cuts first appear non-trivially, and find agreement. As a test of the method, we also applied the same techniques to compute the divergences of the unrenormalized amplitude and found agreement with the sum of Feynman diagrams for subleading divergences up to 4-loop order. The final and most powerful check is consistency with the renormalization group to all loop orders, which we check in the next section. The details of the recursion relations, including subtleties of the phase-space integration, appear in \cite{CMMS}.

\vspace{-1em}
\section{Renormalization}
Recall that the CS equation is given by \autoref{eq:CSonM}. The $\beta$ and $\gamma$ functions themselves admit a Taylor expansion in $\lambda$:
\begin{equation}
\beta\left(\lambda\right) = \sum_{L = -1}^{\infty} \frac{\beta_L \lambda^{L + 1}}{\left(16\pi^2\right)^L}, \quad
\gamma\left(\lambda\right) = \sum_{L = 0}^{\infty} \frac{\gamma_L \lambda^L}{\left(16\pi^2\right)^L}. 
\end{equation}
We choose the summation index and normalization to match common conventions. Plugging the ansatze \autoref{eq:Ansatz4} and \autoref{eq:Ansatz2} into the CS equation immediately gives $\beta_{-1} = \beta_0 = \gamma_0 = 0$. On the other hand, using the leading and subleading recursion relations obtained from unitarity, we find 
%that the one-loop anomalous dimension $\gamma_1$ is $0$ while
\begin{equation}
\label{eq:betaandgamma}
\beta_1 = 3, \qquad \beta_2 = -\frac{17}{3}, \qquad \gamma_1 = 0 , \qquad \gamma_2 = \frac{1}{12}.
\end{equation}
Note that, despite its appearance in \autoref{eq:SLL4}, $m_{1,0}$ canceled out in the final expression for $\beta_2$.
This means that the $\beta$ and $\gamma$ functions are scheme-independent through two loops, which is a known result, see, e.g., \cite{Arnone:2003pa}.
We verified \autoref{eq:betaandgamma} by matching to the $\beta$ and $\gamma$ functions in \cite[Sec. 10.3]{Kleinert:2001} computed in the $\overline{\text{MS}}$ scheme.

\textbf{Comparison with the RG recursion.} While this is a sufficient verification of consistency, it is interesting to compare the form of the recursion relations obtained from unitarity to those obtained from the CS equation. These are obtained by plugging the ansatze into \autoref{eq:CSonM} and matching in powers of $\lambda$ and log. The resulting leading recursion from the CS equation is now
\begin{equation}
m_{L, L} = \frac{\beta_1}{2} m_{L - 1, L - 1}.
\end{equation}
At one-loop order, the CS equation also constrains $\beta_1 = 6m_{1, 1}$. With this information, we find
\begin{equation}
m_{L, L} = \frac{1}{3}\left(3m_{1, 1}\right)^L.
\end{equation}
It is interesting to note that the boundary condition for this recursion relation is given by the one-loop result as opposed to the tree-level result as in unitarity. Since $m_{1, 1} = 1/2$ by explicit computation, this matches \autoref{eq:LL4}. This pattern persists to subleading order, where the recursion is
\begin{equation}
\begin{aligned}
m_{L, L - 1} &= \frac{L\beta_1}{2\left(L - 1\right)} m_{L - 1, L - 2} + \frac{\beta_2}{2} m_{L - 2, L - 2} \\
&\hspace{0.4cm} + \frac{2\gamma_2}{L - 1} m_{L - 2, L - 2}.
\end{aligned}
\end{equation}
In this case, in order to determine the subleading coefficient at general $L$-loop order, one must know both the two-loop $\beta$ function, two-loop $\gamma$ function, and $m_{1, 0}$. Alternatively, one could use $m_{1,0}$ and $m_{2, 1}$, since they are related to $\beta_2$ and $\gamma_2$. Similarly, the recursion relation for the 2-particle amplitude is 
\begin{equation}
n_{L, L - 1} = \frac{\beta_1}{2} n_{L - 1, L - 2},
\end{equation}
with $n_{2, 1} = - \gamma_2$. Knowing $\gamma_2$ and $n_{1, 0}$ (or $n_{1, 0}$ and $n_{2, 1}$) then fixes the general $L$-loop result. With the correct values of the initial conditions, the solutions to these recursion relations match \autoref{eq:LL2} and \autoref{eq:SLL4}.

To summarize, while the constraints obtained from unitarity satisfy the constraints obtained from RG, unitarity constraints start at one lower loop order. The initial data needed for the all loop recursion relation in the case of RG are the functions $\beta$ and $\gamma$ as well as the constant amplitude coefficients. Instead, unitarity determines the general $L$ loop coefficient solely using the generically scheme-dependent constant pieces $m_{L, 0}$ and $n_{L, 0}$ as boundary conditions. From this point of view, these constant pieces in the amplitude play a similar role in unitarity to $\beta$ and $\gamma$ in RG. The summary of the role of each of the coefficients in the recursion is given in \autoref{fig:table}. The relations shown in gray would follow from the (strong) UFC.

\begin{figure}[t]
\hspace{-1em}
\begin{tikzpicture}[remember picture]
\matrix (m) [
        matrix of math nodes, % All cell content is in math mode
        nodes={text width=0.6cm, minimum height=0.7cm, anchor=center, align=center}, % Draw a box for each node
        row sep=-\pgflinewidth, % Collapse row borders
        column sep=-\pgflinewidth  % Collapse column borders
    ]
    {
        m_{0,0} & (\beta_1,\gamma_1) &  &  & \\
        |[fill=orange!20]| m_{1,1} &  m_{1,0} & (\beta_2,\gamma_2) &  & \\
        |[fill=orange!20]| m_{2,2} & |[fill=orange!20]| m_{2,1} &  m_{2,0} & (\beta_3,\gamma_3) \\
        |[fill=orange!20]| m_{3,3} & |[fill=orange!20]| m_{3,2} & |[fill=gray!10]| m_{3,1} & m_{3,0} & (\beta_4,\gamma_4) \\
        |[fill=orange!20]| \vdots & |[fill=orange!20]| \vdots & |[fill=gray!10]| \vdots & |[fill=gray!10]| \vdots & \ddots  \\
    };
    % Leading arrows
    \draw [<-, orange!80!black, shorten <=-4pt, shorten >=-4pt] (m-1-1) -- (m-2-1);
    \draw [<-, orange!80!black, shorten <=-4pt, shorten >=-4pt] (m-2-1) -- (m-3-1);
    \draw [<-, orange!80!black, shorten <=-4pt, shorten >=-4pt] (m-3-1) -- (m-4-1);
    \draw [<-, orange!80!black, shorten <=-4pt, shorten >=-4pt] (m-4-1) -- (m-5-1);
    \draw [<-, orange!80!black, shorten <=-4pt, shorten >=-4pt] (m-1-2) -- (m-2-1);
    % Subleading arrows
    \draw [<-, orange!80!black, shorten <=-4pt, shorten >=-4pt] (m-2-2) -- (m-3-2);
    \draw [<-, orange!80!black, shorten <=-4pt, shorten >=-4pt] (m-3-2) -- (m-4-2);
    \draw [<-, orange!80!black, shorten <=-4pt, shorten >=-4pt] (m-4-2) -- (m-5-2);
    \draw [<-, orange!80!black, shorten <=-4pt, shorten >=-4pt] (m-2-3) -- (m-3-2);
    \draw [<-, orange!80!black, shorten <=-4pt, shorten >=-4pt] (m-2-1) -- (m-3-2);
    % Subsubleading arrows
    \draw [<-, gray, shorten <=-4pt, shorten >=-4pt] (m-3-3) -- (m-4-3);
    \draw [<-, gray, shorten <=-4pt, shorten >=-4pt] (m-4-3) -- (m-5-3);
    \draw [<-, gray, shorten <=-4pt, shorten >=-4pt] (m-3-4) -- (m-4-3);
    \draw [<-, gray, shorten <=-4pt, shorten >=-4pt] (m-3-2) -- (m-4-3);
    % Subsubsubleading arrows
    \draw [<-, gray, shorten <=-4pt, shorten >=-4pt] (m-4-4) -- (m-5-4);
    \draw [<-, gray, shorten <=-4pt, shorten >=-4pt] (m-4-5) -- (m-5-4);
    \draw [<-, gray, shorten <=-4pt, shorten >=-4pt] (m-4-3) -- (m-5-4);
\end{tikzpicture}
\,
\begin{tikzpicture}[remember picture]
\matrix (n) [
        matrix of math nodes, % All cell content is in math mode
        nodes={text width=0.6cm, minimum height=0.7cm, anchor=center, align=center}, % Draw a box for each node
        row sep=-\pgflinewidth, % Collapse row borders
        column sep=-\pgflinewidth  % Collapse column borders
    ]
    {
        n_{0,0} & (\beta_1,\gamma_1) &  &  & \\
        |[fill=orange!20]| n_{1,1} &  n_{1,0} & (\beta_2,\gamma_2) &  & \\
        |[fill=orange!20]| n_{2,2} & |[fill=orange!20]| n_{2,1} &  n_{2,0} & (\beta_3, \gamma_3) \\
        |[fill=orange!20]| n_{3,3} & |[fill=orange!20]| n_{3,2} & |[fill=gray!10]| n_{3,1} & n_{3,0} & (\beta_4, \gamma_4) \\
        |[fill=orange!20]| \vdots & |[fill=orange!20]| \vdots & |[fill=gray!10]| \vdots & |[fill=gray!10]| \vdots & \ddots  \\
    };
    % Leading arrows
    \draw [<-, orange!80!black, shorten <=-4pt, shorten >=-4pt] (n-1-1) -- (n-2-1);
    \draw [<-, orange!80!black, shorten <=-4pt, shorten >=-4pt] (n-2-1) -- (n-3-1);
    \draw [<-, orange!80!black, shorten <=-4pt, shorten >=-4pt] (n-3-1) -- (n-4-1);
    \draw [<-, orange!80!black, shorten <=-4pt, shorten >=-4pt] (n-4-1) -- (n-5-1);
    \draw [<-, orange!80!black, shorten <=-4pt, shorten >=-4pt] (n-1-2) -- (n-2-1);
    \draw [<-, orange!80!black, shorten <=-4pt, shorten >=-4pt] (n-1-2) -- (n-2-1);
    % Subleading arrows
    \draw [<-, orange!80!black, shorten <=-4pt, shorten >=-4pt] (n-2-2) -- (n-3-2);
    \draw [<-, orange!80!black, shorten <=-4pt, shorten >=-4pt] (n-3-2) -- (n-4-2);
    \draw [<-, orange!80!black, shorten <=-4pt, shorten >=-4pt] (n-4-2) -- (n-5-2);
    \draw [<-, orange!80!black, shorten <=-4pt, shorten >=-4pt] (n-2-3) -- (n-3-2);
    \draw [<-, orange!80!black, shorten <=-4pt, shorten >=-4pt] (n-2-1) -- (n-3-2);
    % Subsubleading arrows
    \draw [<-, gray, shorten <=-4pt, shorten >=-4pt] (n-3-3) -- (n-4-3);
    \draw [<-, gray, shorten <=-4pt, shorten >=-4pt] (n-4-3) -- (n-5-3);
    \draw [<-, gray, shorten <=-4pt, shorten >=-4pt] (n-3-4) -- (n-4-3);
    \draw [<-, gray, shorten <=-4pt, shorten >=-4pt] (n-3-2) -- (n-4-3);
    % Subsubsubleading arrows
    \draw [<-, gray, shorten <=-4pt, shorten >=-4pt] (n-4-4) -- (n-5-4);
    \draw [<-, gray, shorten <=-4pt, shorten >=-4pt] (n-4-5) -- (n-5-4);
    \draw [<-, gray, shorten <=-4pt, shorten >=-4pt] (n-4-3) -- (n-5-4);
\end{tikzpicture}

\begin{tikzpicture}[remember picture, overlay]
\draw [->, orange!80!black, shorten <=-4pt, shorten >=-4pt] ([xshift=15pt]m-1-2.east) -- ([xshift=-0.1cm]n-1-1.west);
\end{tikzpicture}

    \caption{Summary of the results. Coefficients of the 4-particle (left) and 2-particle (right) amplitudes $m_{L,k}$ and $n_{L,k}$ can be fixed either through unitarity (sourced by $m_{L,0}$ and $n_{L,0}$) or RG (sourced by $\beta_L$ and $\gamma_L$ as well as $m_{L, 0}$ and $n_{L, 0}$). The connecting line indicates that the 2-particle amplitude required the 4-particle amplitude. Although the initial conditions are different, the solutions to these recursions are equivalent. Scheme dependence is generically encoded both in the constant terms $m_{L, 0}$ and $n_{L, 0}$ as well as the coefficients $\beta_L$ and $\gamma_L$. Orange shading denotes proven unitarity relations while those in gray are conjectured by the UFC.}
    \label{fig:table}
\end{figure}

\vspace{-1em}
\section{Discussion}
\vspace{-0.5em}

We established, in the massless $\lambda \phi^4$ theory, a concrete equivalence between the constraints imposed by unitarity and those encoded in the renormalization group equation, thus proving the strong UFC in this theory through the subleading order.
Given the assumptions made, several extensions are possible, in which we hope at least a weak form of the UFC could hold:

\textbf{Mass renormalization.} The techniques developed in this paper, when applied to a massive theory, will still be able to reproduce the $\beta$ and $\gamma$ functions through unitarity cuts of massive propagators. However, unitarity cuts cannot access massive tadpole diagrams and are therefore insensitive to mass renormalization. Generalized cuts \cite{Bern:1994cg} of massive tadpoles may give a way to incorporate mass renormalization into our analysis.

\textbf{IR divergences/rapidity logarithms.} In forward/Regge limits, an additional scale must be introduced that runs between soft and collinear modes, which then appear in rapidity logarithms \cite{Chiu:2011, Rothstein:2023dgb}. A natural site to explore whether the UFC would apply in these situations is soft collinear effective theory, inspired by results in \cite{Rothstein:2024}.  

\textbf{Operator mixing.} Including a tower of higher dimension operators leads to intricate mixing during renormalization due to the introduction of a new coupling at each order in EFT power counting. Work on computing anomalous dimensions with on-shell techniques in the presence of operator mixing has been done, for example, in \cite{Bresciani:2023}. Our methods, which rely on the non-perturbative optical theorem and avoid counterterms, are organized by particle cuts rather than loop order and so may streamline such computations.

\textbf{Bootstrap considerations.} 
One may view our results as a version of the $S$-matrix bootstrap in which aspects of the RG flow have been derived using physical axioms of unitarity, symmetry, and analyticity. It is worth exploring to what extent this strategy can be combined with the established non-perturbative bootstrap \cite{Kruczenski:2022lot} and positivity \cite{deRham:2022hpx} programs. In this context, it would be necessary to understand the effect of non-analyticities. For example, one could explore the interplay between the monodromy group of the S-matrix (capturing analyticity on different Riemann sheets) and the renormalization group.

\paragraph*{\bf Acknowledgments.}
We thank Nima Arkani-Hamed, Miguel Correia, Frederik Denef, Mathieu Giroux, and Giulio Salvatori for useful comments and discussions. A.C. is supported by the National Science Foundation Graduate Research Fellowship under Grant No. DGE-2036197. The work of D.M. and J.S. is supported by the US Department of Energy grant DE-SC011941.

\bibliography{references1}
\bibliographystyle{JHEP}

\end{document}